# Consistency of time averaged optical force laws for embedded chiral and achiral objects

M.R.C. Mahdy[1,2†*], Tianhang Zhang[1,3†], Weiqiang Ding[4*], Amin Kianinejad[1], Manuel Nieto-Vesperinas[5]

[1]*Department of Electrical and Computer Engineering, National University of Singapore, 4 Engineering Drive 3, Singapore 117583*

[2]*Department of Electrical & Computer Engineering, North South University, Bashundhara, Dhaka 1229, Bangladesh*

[3]*NUS Graduate School for Integrative Sciences and Engineering, National University of Singapore, 28 Medical Drive, Singapore 117456*

[4]*Department of Physics, Harbin Institute of Technology, Harbin 150001, People's Republic of China*

[5]*Instituto de Ciencia de Materiales de Madrid, Consejo Superior de Investigaciones Cientificas, Campus de Cantoblanco, 28049, Madrid, Spain*

[*] Corresponding authors: A0107276@u.nus.edu and wqding@hit.edu.cn

[†]These Authors contributed equally to this work.

**Although it is commonly believed that all the volumetric optical force laws should lead to the same total optical force for chiral and achiral objects, this idea has been invalidated in some recent works by investigating several previous experiments involving material background. To identify the exact reason of such significant disagreement, we inspect two tractor beam and one lateral force experiments on using distinct stress tensors (STs). To solve the problems of total force, we propose two 'consistency conditions' of time averaged forces. We demonstrate that exactly at the boundary of an object, the difference of the consistent external Minkowski ST and internal ST of Chu (and Einstein-Laub) is found in agreement with the surface force yielded by Chu (and Einstein-Laub) force only when the background is air rather than a material. We identify this as one of the main reasons (among few other identified reasons) of the disagreements observed for real experiments. Finally, based on the proposed 'consistency conditions', we demonstrate that: by modifying the Einstein-Laub or Chu formulation, time-averaged STs and volume forces are obtainable those can overcome the aforementioned inconsistencies of real experiments for both chiral and achiral Mie objects embedded in even complex material backgrounds.**

**INTRODUCTION**

A tractor beam is a customized light beam which exerts a negative force on a scatterer [1-11], pulling it opposite to the propagation direction of light, in contrast to the conventional pushing forces. Not only the counterintuitive idea of optical pulling (i.e. the tractor beam effect) but also the idea of optical lateral force [12-15] is also growing well-known in recent literature. Tractor beam and lateral force experiments, which involve the material background [1, 2, 9, 11, 12], can also be investigated in details to understand the persistently debated roles of different STs, forces and photon momenta such as those of the Abraham-Minkowski controversy [16-19]; as the employment of inappropriate approach may lead to pushing force (or inconsistent lateral force) instead of the experimentally observed pulling one (or consistent lateral force). Throughout this paper we refer to 'exterior' or 'outside' magnitudes those evaluated outside the volume of the embedded macroscopic object, while by 'interior' or 'inside' we shall refer to those quantities inside this object volume [cf. the force calculation process in Fig.1 (a) - (c)]. Although it is commonly believed all the volume force laws [18, 20] [i.e. Minkowski, Abraham, Chu, Einstein-Laub (EL) and Ampere/Nelson] should lead to the same time averaged total optical force, this idea has been invalidated in recent works [3, 19] considering several experimental set-ups [19] involving material background. In fact, if one involves the material background; several complexities arise:

(i) In order to calculate the total force on an embedded object, normally two conventions, i.e., GAP METHOD AND NO GAP METHOD, are applied [20-25] but which one is more accurate than the other is still not properly understood. (a) GAP METHOD [20, 23-25]: By introducing a very small gap between the scatterer and the background to yield the force as shown in Fig. 1(a). (b) NO GAP METHOD [1-3,12,19,21,22]: Without introducing any small gap outside the embedded scatterer as shown in Fig. 1(b) and (c) and in supplement S1. Between these two distinct methods, which one is more appropriate?

(ii) In addition, the involvement of chirality [26-36] can be a key factor to judge the consistency of the optical force calculation methods which are considered applicable for embedded achiral objects. Previous analyses of optical force in chiral objects are still restricted by the following factors: (a) the dipolar approximation to the scattering object [13, 29, 30, 32], (b) the background being just air or vacuum [29-31, 33-36] and (c) the internal force distribution inside a slab embedded in air [35,36]. But the volumetric force distribution formula, along with its stress tensor for any arbitrary generic embedded chiral object [26, 27, 32], used in real experiments [26, 27] are still not investigated in details.

This work attempts to solve the aforementioned complexities and investigates two tractor beam and one lateral force experiments. We demonstrate that even without introducing any artificial gap between the background and the embedded object, it is possible to establish different equivalent time-averaged force formulas based on the fulfilment of just two 'consistency conditions' [denoted as C(I) and C(II) with details shown below]. This may also solve the reported problems [2, 3, 19] of total force calculation for almost all real experiments conducted so far involving a material background. In addition, this work introduces the less time and memory-consuming internal Modified Einstein-Laub (MEL) stress tensor method, (rather than the longer time-consuming bulk volumetric force method [21, 39]) for both achiral [1-3, 19] and chiral objects [26, 27, 32] embedded in a generic (i.e. homogeneous, heterogeneous, bounded etc.) background.

**PROPOSAL OF THE TWO 'CONSISTENCY CONDITIONS'**

One of the fundamental proposals of this paper is: if $\bar{\bar{T}}(\text{in})$ [applied at $r=a^-$ employing only the interior field of a scatterer, where $a$ is the particle radius; cf. Fig. 1 (b) and (c)] is a valid internal stress tensor and $\boldsymbol{f}(\text{in})$ is its corresponding volume force; and $\bar{\bar{T}}(\text{out})$ is a valid external stress tensor which is applied at $r=a^+$ using only exterior fields of an object; then the following 'validity condition' should be fulfilled:

"On the boundary $r=a$ of any object, this equation: $[\bar{\bar{T}}(\text{out}) - \bar{\bar{T}}(\text{in})] \cdot \hat{\boldsymbol{n}} = \boldsymbol{f}^{\text{Surface}}$ should hold ($\hat{\boldsymbol{n}}$ being the local unit outward normal of the object surface). Also, the same surface force $\boldsymbol{f}^{\text{Surface}}$ should independently be found from the volume force density $\boldsymbol{f}(\text{in})$ by applying the appropriate boundary conditions at $r=a$. These two aforementioned conditions must be satisfied simultaneously."

For example- in [37], the process of obtaining $f^{Surface}$ has been shown only from a volumetric bulk force density equation, $f$ (in), stemming from the well-known Chu and Einstein-Laub force, and by considering only air as the background. This article also attempts the same approach for volumetric force ($f$) but an additional approach with STs ($[\bar{\bar{T}}(\text{out}) - \bar{\bar{T}}(\text{in})] \cdot \hat{n} = f^{Surface}$) due to the complexity of the presence of material background.

At the beginning of supplement S2, it is shown that when the background is air, the difference of the external ST (notice that then all STs are same) and the internal ST of Chu (and of Einstein-Laub) at the object boundary is in complete agreement with the fully independently calculated surface force given by the volumetric formulation of Chu and Einstein-Laub. But when material background is involved: the first question should be: Q(I) which $\bar{\bar{T}}(\text{out})$ should be appropriate? Then the second question is: Q(II) If $\langle F^{Bulk} \rangle (\text{in}) = \int \langle f^{Bulk} \rangle (\text{in}) \cdot dv = \int \langle \bar{\bar{T}}(\text{in}) \rangle \cdot ds \neq 0$ [as bulk force represents the volumetric force distribution and it cannot be ignored for several situations]; which $\bar{\bar{T}}(\text{in})$, along with its corresponding $f$, will satisfy these two following 'consistency conditions' simultaneously: C(I) the aforementioned 'validity condition' and C(II) the consistent time-averaged force equation:

$$\int \langle \bar{\bar{T}}(\text{out}) \rangle \cdot ds = \langle F^{Total}(\text{Consistent}) \rangle = \langle F^{Bulk} \rangle (\text{in}) + \langle F^{Surface} \rangle ?$$

In this article we show that when the background is a material medium instead of air, $\bar{\bar{T}}(\text{out})$ and $\bar{\bar{T}}(\text{in})$ cannot be arbitrary STs which satisfy C(I) and C(II) simultaneously.

**WHAT IS DEMONSTRATED IN THIS ARTICLE**

Concerning Q(I) in previous section, in order to sort out the appropriate ST $\bar{\bar{T}}(\text{out})$, we investigate two major tractor beam experiments [1, 2] along with the recent set-up of a lateral force experiment [12]. We identify that the vacuum ST [20] of the GAP method leads to *inconsistent results* for tractor beams and lateral force experiments, especially when the symmetry is broken [2,12]. Both external Minkowski and Abraham STs of a NO GAP METHOD lead to a consistent time-averaged total force for all those experiments, (and also for all previous experiments [19,38]). However, this does not mean external STs other than Minkowski (or Abraham) are incorrect. We shall investigate this issue further throughout the article. But an important point to note that:

'*even to sort out appropriate $\bar{\bar{T}}(\text{out})$, it is required to satisfy aforementioned consistency conditions C(I) and C(II) simultaneously.*'

Now, we shall consider Q(II) of previous section. As always $\langle F^{\text{Bulk}} \rangle (\text{in}) = \int \langle f^{\text{Bulk}} \rangle (\text{in}) \cdot dv = \int \langle \bar{\bar{T}}(\text{in}) \rangle \cdot ds = 0$ for internal Minkowski [or Helmholtz force] and internal Abraham ST [or Abraham force] for the transparent/non-absorbing objects, to sort out the appropriate $\bar{\bar{T}}(\text{in})$ [and also $\bar{\bar{T}}(\text{out})$ of Q(I)] we examine several cases for an object embedded in material background [2,3,19, 26, 27, 32, 38-45].

For example- condition C(I) has been violated when: (i) $\bar{\bar{T}}(\text{out})$ is considered $\bar{\bar{T}}_{\text{Mink.}}(\text{out})$ and $\bar{\bar{T}}(\text{in})$ is well known internal Chu or Einstein-Laub ST; and (ii) both $\bar{\bar{T}}(\text{out})$ and $\bar{\bar{T}}(\text{in})$ are considered $\bar{\bar{T}}_{\text{EL}}$ respectively. More details will be discussed later. In contrast, C(II) is seriously violated when: (iii) both $\bar{\bar{T}}(\text{out})$ and $\bar{\bar{T}}(\text{in})$ are considered $\bar{\bar{T}}_{\text{Chu}}$ respectively. More details will be discussed later. (iv) We have already discussed very shortly that external vacuum ST of GAP METHOD violates the left-hand side equation of C(II). These are identified as the main reasons of disagreements observed in [19] and [38] for different volumetric force laws.

Finally we demonstrate that when $\langle F^{\text{Bulk}} \rangle (\text{in}) \neq 0$ for non-absorbing objects; to satisfy C(I) and C(II) simultaneously for a chiral or achiral Mie scatterer embedded in generic material backgrounds [i.e. homogeneous, heterogeneous, bounded or unbounded], $\bar{\bar{T}}(\text{out})$ should be considered as Minkowski ST and $\bar{\bar{T}}(\text{in})$ should be the modified version of Einstein-Laub (or Chu) formulation along with modified *f*(in). So, when $\langle F^{\text{Bulk}} \rangle (\text{in}) \neq 0$, in order to satisfy both C(I) and C(II) simultaneously, $\bar{\bar{T}}(\text{out})$ and $\bar{\bar{T}}(\text{in})$ cannot have the same form for embedded transparent scatterers. To explain the last observation for realistic situations [2,3,19,26,27,32,38-45], we conclude at the end: though both the external Minkowski (and Abraham) ST and proposed internal MEL (and Modified Chu) methods lead to the same consistent time-averaged total force [i.e. they are mathematically equivalent], they should better be considered two fully different operations/process from the physical point of view.

**RESULTS AND DISCUSSIONS**

**The consistent external force with NO GAP METHOD**

First we investigate the consistency of the total force calculation for one tractor beam experiment [1] and for one on a lateral force [12], using the major stress tensors, namely: Minkowski, Chu, Ampere and Einstein-Laub with NO GAP METHOD [cf. Fig. 1(b) and (c)]. All 3D simulations throughout the paper are conducted using an incident power of 0.57mW/μm$^2$. For example, the external time-averaged total outside force from Minkowski's ST is written as [1-3,12]:

$$\langle \boldsymbol{F}_{\text{Total}} \rangle (\text{out}) = \int \langle \overline{\overline{\boldsymbol{T}}}_{\text{Mink}}^{\text{out}} \rangle \cdot d\boldsymbol{s}, \tag{1a}$$

$$\langle \overline{\overline{\boldsymbol{T}}}_{\text{Mink}}^{\text{out}} \rangle = \frac{1}{2}\text{Re}\left[ \boldsymbol{D}_{\text{out}} \boldsymbol{E}_{\text{out}}^* + \boldsymbol{B}_{\text{out}} \boldsymbol{H}_{\text{out}}^* - \frac{1}{2}\left( \boldsymbol{B}_{\text{out}} \cdot \boldsymbol{H}_{\text{out}}^* + \boldsymbol{D}_{\text{out}} \cdot \boldsymbol{E}_{\text{out}}^* \right) \overline{\overline{\boldsymbol{I}}} \right]. \tag{1b}$$

'out' stands for fields outside the object, [e.g., on $r=a^+$, if it is a sphere or cylinder of radius $a$, cf. Fig. 1(b)] and $\overline{\overline{\boldsymbol{I}}}$ is the unity tensor. The electromagnetic vectors in Eq. (1b) correspond to the total field, namely, incident plus field scattered by the body. For the force calculations with other stress tensors (i.e. Chu, Ampere and Einstein-Laub), we shall also use this 'total' outside fields.

Without introducing any small gap (named as NO GAP METHOD in the introduction), total force has been calculated by different external stress tensors for two beam tractor configuration reported in [1], in supplement S1 Fig. 1s(a) - (c) and also for lateral force experiment [12] configuration in Fig. 2(a) - (c). Though Minkowski's (or Abraham's) formulation leads to the most accurate time averaged force for some other experiments [2, 3, 46-49] reported in [19]; for two tractor beam [1] and lateral force experiments [12] it may not be possible to recognize which one is the most consistent external ST. Due to the very big size of the object [12] considered in the real lateral force experiment [special technique has been applied in the first ref. of [12] due to the big size of particle: 4500 nm], we are modeling that set-up with comparatively small sized object [cf. the second reference in [12], where a small object has been considered]. However, we are considering two cases: spherical and elliptical object to check the consistency of the lateral force considering the arbitrary shape of the testing object. Previously for interfacial tractor beam experiment [2,3], only external Minkowski ST lead to the accurate prediction for arbitrary shaped objects [2,3] modelled as spherical and elliptical shaped objects. Though there is no difference among the signs of the total forces for the experiments reported in [1] and [12], their magnitudes are observed quite different in supplement Figs 1s(b), (c) [two beam tractor] and in main article Figs 2(b), (c) [lateral force]. But the important fact is that: For the NO GAP METHOD: (A) The size based sorting of embedded particles by two beam method [cf. supplement of ref. [1]] is consistently predicted by all external STs. (B) The sign change of the force for two different handedness of polarizations observed for lateral force experiment [12] has also been correctly predicted by all external STs for spherical and elliptical objects respectively. In Fig. 3(a) given in the second ref. of [12], for 1500 nm sized object, a single direction of lateral force has been observed for a single handedness of circularly polarized light (consistent with our observation). These aforementioned observations will be imperative for our next investigation: the consistency of GAP METHOD for the tractor beam experiments [1,2] and the lateral force experiment [12].

**Inconsistency of the GAP METHOD and different other formulations**

One convention of calculating optical force is that if the background is material medium; the force should be calculated considering an extremely small gap [cf. Fig. 1(a)] between the embedded object and the background as discussed in ref. [20,23-25]. As a result, in this section we shall investigate the consistency of this GAP METHOD for the scatterers embedded in material media such as in [1], [2] and [12]. According to [20], if a small gap is introduced [cf. Fig. 1(a)], the external vacuum stress tensor (Chu type stress [20]) to yield the total outside force should be written in terms of gap fields ($E_g, H_g$) as:

$$\left\langle \boldsymbol{F}_{\text{Total}} \right\rangle (\text{out}) = \int \left\langle \bar{\bar{\boldsymbol{T}}}_{\text{Vacuum}}^{\text{out}} \right\rangle \cdot d\boldsymbol{s}, \tag{2a}$$

$$\left\langle \bar{\bar{\boldsymbol{T}}}_{\text{Vacuum}}^{\text{out}} \right\rangle = \frac{1}{2} \text{Re} \left[ \varepsilon_0 \boldsymbol{E}_g \boldsymbol{E}_g^* + \mu_0 \boldsymbol{H}_g \boldsymbol{H}_g^* - \frac{1}{2} \left( \mu_0 \boldsymbol{H}_g \cdot \boldsymbol{H}_g^* + \varepsilon_0 \boldsymbol{E}_g \cdot \boldsymbol{E}_g^* \right) \bar{\bar{\boldsymbol{I}}} \right]. \tag{2b}$$

The gap fields ($E_g$, $H_g$) and time-averaged optical forces in this article have been computed based on both COMSOL MULTIPHYSICS (most of the cases) and Lumerical softwares [39]. The homogeneous background cases of NO GAP METHOD have been calculated both analytically (Mie theory) and numerically (full-wave based computer simulations). In Fig. 3(a) we have obtained the external force considering such an extremely small gap [cf. Fig. 1(a)] between the half immersed scatterer and the water background and by employing the stress tensor of Eq. (2b). Instead of the experimentally observed pulling force [2, 3], we obtain pushing force due to such an extremely small gap as shown in Fig. 3(a). In contrast, both GAP METHOD and NO GAP METHOD lead to consistent pulling force for two beam tractor experiment [1] as shown in Fig. 3(b). Then we have considered the lateral force experiment reported in [12]. Though for the spherical sized object the total force seems in good agreement [cf. Fig. 3(c)] with the previously calculated forces with NO GAP METHOD [cf. Fig. 2 (b)], for elliptical shaped object the sign significantly alters [cf. Fig. 3(d)] in comparison with NO GAP METHOD [cf. Fig. 2(c)]. So, the problem of GAP METHOD mainly arises when the background is inhomogeneous/symmetry broken case [2,3,12] (or in general heterogeneous type as shown in Fig. 1(b)) [serious violation of the left side of C(II)].

Another detail analysis has been done in favor of GAP METHOD in ref. [23] (and also in refs. [24] and [25]). Instead of the external ST Eq. (2b), it is suggested in [23] to yield the total force by an appropriate volume force for some specific cases (described as method II in [23]). However, in supplement S2 we have discussed in details why method II reported in [23] may not be a general way to yield the total force and to explain the so far reported experiments, especially for the experiments with inhomogeneous (or heterogeneous) background specially due to the violation of C(I).

In fact, even considering no such gap, the well-known Einstein-Laub volume force [19] (but not Chu and Amperian/Nelson force) predicts consistent time averaged total force for almost all previous experiments as shown in details in [19]. However, the magnitude of the total force by EL law is not in full/exact agreement with the total force calculated by external Minkowski ST or Helmholtz force [19,38] for *all those experiments*. Regarding the consistency of the EL force for the case of a material background, one has first to examine C(I) and C(II). This issue is discussed next.

In supplement S2 it is shown that exactly at the object boundary, the difference of well-established external Minkowski ST and internal Chu ST (also similarly applicable for Einstein-Laub ST) is found in agreement with the surface force of Chu (or Einstein-Laub) when the background is air. But if the background is any material medium instead of air or vacuum, this conclusion does not remain true. For example, condition C(I) has been violated when: (i) $\bar{\bar{T}}(\text{out})$ is considered $\bar{\bar{T}}_{\text{Mink.}}(\text{out})$ and $\bar{\bar{T}}(\text{in})$ is well known internal Chu or Einstein-Laub ST; and (ii) $\bar{\bar{T}}(\text{out})$ and $\bar{\bar{T}}(\text{in})$ both are considered $\bar{\bar{T}}_{\text{EL}}$ respectively. It should also be noted that EL force leads to inconsistent result for few experiments [violation of C(II)]: Hakim-Higham experiment [38], Rasetti experiment [50], few cases of Jones' experiments according to [19] [cf. Fig. 9 (a)-(d) given in [19]] and some other notable cases reported in [51].

By contrast, C(II) is seriously violated when: (iii) $\bar{\bar{T}}(\text{out})$ and $\bar{\bar{T}}(\text{in})$ both are considered $\bar{\bar{T}}_{\text{Chu}}$ respectively. For example, when $\langle F^{\text{Total}} \rangle = \langle F^{\text{Bulk}} \rangle (\text{in}) + \langle F^{\text{Surface}} \rangle$ is calculated by employing the internal volumetric force of Chu [or by Chu ST: $\langle F^{\text{Bulk}} \rangle (\text{in}) = \int \langle \bar{\bar{T}}_{\text{Chu}}(\text{in}) \rangle \cdot ds$] for the several real experiments reported in [19], it does not lead to the correct time averaged total force, $\langle F^{\text{Total}}(\text{Consistent}) \rangle$ [Violation of right hand side of the equation given in C(II)]. In addition, in [3] it is shown that left-hand side of equation of C(II) has been violated for $\bar{\bar{T}}_{\text{Chu}}(\text{out})$ based on NO GAP METHOD.

- All of these are identified as the main reasons of disagreements observed in [19] (and also in [38]) for different volumetric forces.

In the next section we shall demonstrate that not only the surface force but also the bulk force of the well-known Einstein-Laub (or Chu) force law is responsible for such disagreements those reported in [19] and [38] for the real experiments. It would be possible to overcome such inconsistencies if and only if the GAP METHOD were applicable where the external Minkowski ST turns into Vacuum ST [cf. Eq. (2b)]. But in this section we have already demonstrated the problem/inconsistency of the GAP

METHOD. Hence the possible solution of such problems will be addressed in the next three sections.

**Consistency of the external Minkowski and internal MEL or modified Chu formulations**

Though so far Minkowski's ST has led to consistent time-averaged results for real experiments [19,38] by employing the exterior field of an embedded scatterer, [fulfilment of the left side of the equation in C(II)], more complex configurations may arise where Minkowski's ST and its associated Helmholtz force would fail [i.e. cf. refs. [52-54]]. Hence we shall now examine whether or not the external Minkowski ST leads to the fulfilment of both C(I) and C(II) together with any other internal ST and force. To handle future situations, and to put forward new optical force laws if necessary, the validity conditions C(I) and C(II) together and the analysis that follows below should be effective.

A modification of the well-known EL volume force (and also of Chu's force [21]) has been predicted in [22]. That modified version of the EL force led to exactly the same time-averaged force predicted by the external Minkowski ST in [22] for a dielectric object embedded in another dielectric. Commercial software [39] applies this modified version of the Einstein-Laub [22] or the Chu volume force density [21] to yield the total volumetric force on dielectric objects embedded in another dielectric. On the other hand, in order to explain the interfacial tractor beam experiment [2], the achiral internal MEL ST [3] has been applied previously without any derivation and consistent description. So far no explanation is given in [3] and [22] why both Minkowski ST and the MEL volume force [22] (and also modified Chu force [21,39]) and the MEL ST [3] lead to same/consistent time averaged result for dielectric cases. This section (and the final conclusion of this paper) explains the reason, even considering more general cases such as magnetodielectric objects embedded in a generic magnetodielectric background.

The achiral MEL ST (which should yield the bulk force of an embedded scatterer) inside [at $r=a^-$] an object embedded in a generic heterogeneous background should be written as:

$$\left\langle \overline{\overline{T}}_{\text{MEL}(j)}^{\text{Bulk}} \right\rangle (\text{in}) = \boldsymbol{D}_{\text{in}} \boldsymbol{E}_{\text{in}}^* + \boldsymbol{B}_{\text{in}} \boldsymbol{H}_{\text{in}}^* - \frac{1}{2}\left( (\frac{\mu_{b(j)}}{\mu_s})\mu_s \boldsymbol{H}_{\text{in}}^* \cdot \boldsymbol{H}_{\text{in}} + (\frac{\varepsilon_{b(j)}}{\varepsilon_s})\varepsilon_s \boldsymbol{E}_{\text{in}}^* \cdot \boldsymbol{E}_{\text{in}} \right) \overline{\overline{I}}. \qquad (3)$$

where $j=1,2,3,...., N$ represents the number of background regions sharing interface with the object, [cf. our Figs. 1(b), (c) and forthcoming Fig. 5(a)]. $\varepsilon_b$ and $\mu_b$ are fixed background permittivity and permeability and $\varepsilon_s$ and $\mu_s$ are fixed permittivity and permeability of the scatterer. A possible derivation of Eq. (3) is given in Supplement S3 where we have shown that the achiral MEL ST should be written in the time averaged form along with its time averaged volume force density (i.e. MEL force law for magnetodielectric object embedded in a generic magnetodielectric heterogeneous

background):

$$\langle \boldsymbol{f}_{\mathrm{MEL}(j)}^{\mathrm{Bulk}} \rangle (\mathrm{in}) = \frac{1}{2} \mathrm{Re}\left[ \left(\boldsymbol{P}_{\mathrm{Eff}(j)} \cdot \nabla\right) \boldsymbol{E}_{\mathrm{in}}^* + \left(\boldsymbol{M}_{\mathrm{Eff}(j)} \cdot \nabla\right) \boldsymbol{H}_{\mathrm{in}}^* - \left(i\omega \boldsymbol{P}_{\mathrm{Eff}(j)} \times \boldsymbol{B}_{\mathrm{in}}^*\right) + \left(i\omega \boldsymbol{M}_{\mathrm{Eff}(j)} \times \boldsymbol{D}_{\mathrm{in}}^*\right) \right]. \quad (4)$$

A similar formulation can also be written for modified Chu formulation for achiral embedded objects given in [21, 39]. Consistency of Eq. (3) and (4) has been shown for surface force calculation in supplement S4, which satisfies the 'validity condition' [fulfilment of C(I)]. In Eq. (4), the effective polarization [21,22] and magnetization [21] are defined as: $\boldsymbol{P}_{\mathrm{Eff}} = (\varepsilon_S - \varepsilon_b) \boldsymbol{E}_{\mathrm{in}}$ and $\boldsymbol{M}_{\mathrm{Eff}} = (\mu_S - \mu_b) \boldsymbol{H}_{\mathrm{in}}$. The total time averaged Bulk force on the embedded object should be:

$$\langle \boldsymbol{F}_{\mathrm{MEL}}^{\mathrm{Bulk}} \rangle (\mathrm{in}) = \sum_j \int \langle \overline{\overline{\boldsymbol{T}}}_{\mathrm{MEL}(j)}^{\mathrm{Bulk}} \rangle (\mathrm{in}) \cdot d\boldsymbol{s}_{(j)} = \sum_j \int \langle \boldsymbol{f}_{\mathrm{MEL}(j)}^{\mathrm{Bulk}} \rangle (\mathrm{in}) \cdot dv_{(j)}. \quad (5)$$

Consistency of Eq (5) can be verified by considering a magnetodielectric slab embedded in another magnetodielectric. Based on both internal stress tensor and volumetric force process of Eq (5), we arrive at the time averaged force per unit area as: $\frac{1}{2} \frac{E_0^2}{\mu_s \left(\frac{\mu_b \varepsilon_s}{\mu_s \varepsilon_b} - 1\right)} \left[ (\varepsilon_s - \varepsilon_b)\mu_s - (\mu_s - \mu_b)\varepsilon_s \right] \left\{ 1 + |R|^2 - |T|^2 \right\}$. Here $R$ and $T$ denote the reflection and transmission coefficients of the slab and $E_0$ is the amplitude of the plane wave. As the surface force and static part of MEL (or modified Chu method) vanish for a slab illuminated by a plane wave, Eq (5) should fulfill condition C(II). As a result, we now investigate the 'outside force'. Based on external Minkowski ST approach, the final time averaged total 'outside force' per unit area on the aforementioned embedded slab can be written as: $\left[ N_i \hbar k_i + N_r \hbar k_r - N_t \hbar k_t \right]$ (also cf. Eq (60) given in ref. [55] and the slab example in [56] for a simpler case: slab placed in air). $N = \langle S \rangle / (\hbar \omega)$ denotes the external photon flux, $k$ is the wavenumber and $\langle S \rangle$ is the time-averaged Poynting vector. Here '$i$', '$r$' and '$t$' mean incident, reflected and transmitted, respectively. According to our analytical calculation, this last equation leads to the same total force calculated previously by Eq (5). So, C(II) is fully satisfied for this special example by external Minkowski force and internal MEL or modified Chu force. It should be noted that the well-known internal Einstein-Laub, Chu and Ampere/Nelson forces do not lead to this exact agreement even for this simple example of an embedded magnetodielectric slab.

However, still one final question remains: is the condition C(II) satisfied when the surface force of MEL method takes place (i.e. for generic cases)?

To answer this question, now, by applying the proper boundary conditions (cf. supplement S4), the surface force of modified Einstein-Laub method can be written from two different ways: (i) By the volume force method of Eq. (4) [as shown in [37] only for volumetric force] and (ii) from the difference of external Minkowski ST and internal MEL ST in Eq. (3) just at the boundary. These two different ways lead to exactly same result (cf. supplement S4) [fulfilment of C(I)]:

$$f_{MEL}^{Surface} = [\bar{\bar{T}}_{Mink}(out) - \bar{\bar{T}}_{MEL(j)}(in)] \cdot \hat{n}_{r=a}$$
$$= \{\epsilon_{b(j)}(E_{out} - E_{in}) \cdot \hat{n}\}\left(\frac{E_{out} - E_{in}}{2}\right)_{at\ r=a} + \{\mu_{b(j)}(H_{out} - H_{in}) \cdot \hat{n}\}\left(\frac{H_{out} - H_{in}}{2}\right)_{at\ r=a} \quad (6)$$

Eq. (6) explains why in [22], the time averaged result of total force predicted by Minkowski ST is in exact agreement with the MEL volume force reported in [22]. Eq. (6) finally suggests that the total time averaged force calculation by MEL force is indeed equivalent with the total force calculation based on external Minkowski ST or Helmholtz force for any generic case. So, the total time averaged force on an embedded generic object according to MEL method should finally be written as (also applicable for modified Chu formulation [21, 39]):

$$\int \left\langle \bar{\bar{T}}_{Mink}^{out} \right\rangle \cdot ds = \left\langle F_{MEL}^{Bulk} \right\rangle(in) + \left\langle F_{MEL}^{Surface} \right\rangle \quad (7)$$

Hence analytically we have arrived at the fulfillment of both C(I) and C(II). Next, the validity of Eq. (7) [for our proposed C(II)] will be investigated mainly based on numerically [i.e. full wave simulations] calculated results for several complicated examples.

**A short discussion on previous tractor beam and lateral force experiments**

The most probable reason of the success of Minkowski's theory lies on the fact that, in contrast with other force formulations, the external Minkowski ST, the Helmholtz force [2, 3, 19, 38, 57] and the ray tracing method based on the Minkowski photon momentum [2,3] correctly account for the linear increase of transferred/emitted photon momentum at the boundary between the embedded object and the background [2, 3]. Interestingly, when $[1-(\frac{\varepsilon_{b(j)}}{\varepsilon_s})]^2$ is very small for a dielectric object embedded in another dielectric (for example- most of the real experiments [1,2,12]), the surface force of MEL in Eq. (6) almost vanishes, as $\left\langle f_{MEL}^{Surface} \right\rangle \propto [1-(\frac{\varepsilon_{b(j)}}{\varepsilon_s})]^2$. Only for such special cases $\left\langle F_{Mink.}^{Total} \right\rangle(out) \approx \left\langle F_{MEL}^{Bulk} \right\rangle(in)$ [as shown in [3] without any explanation]. In [3], the time averaged external force by Minkowski ST and the internal time averaged bulk force by MEL ST match well because the surface force almost vanishes. But in general, for the exact total force formulation, that surface force should be added with the time averaged bulk force calculated by internal MEL ST [cf. Eq. (7)] so that $\langle \boldsymbol{F}_{\text{Mink}}^{\text{Total}} \rangle (\text{out}) = \langle \boldsymbol{F}_{\text{MEL}}^{\text{Bulk}} \rangle (\text{in}) + \langle \boldsymbol{F}_{\text{MEL}}^{\text{Surface}} \rangle$. If the internal force for the tractor beam experiment in [1, 2] and the lateral force in [12] is calculated, our aforementioned conclusions still remain valid. It is shown in Fig.4 (a) and Fig. 4(b), total external time averaged force by Minkowski ST and the total internal bulk force by MEL ST are in almost full agreement for two beam tractor beam experiment in [1] due to very small value of $[1-(\frac{\varepsilon_{b(j)}}{\varepsilon_s})]^2$. However, the difference between the bulk force of MEL ST and the total force of external Minkowski ST is clearly observable for the lateral force experiment when $[1-(\frac{\varepsilon_{b(j)}}{\varepsilon_s})]^2$ is not very small [i.e. by considering a TiO$_2$ object embedded in air-water interface in Fig. 4(c)]. Also there is an effect of the size/shape change of the object on the bulk force of MEL ST as shown in Fig. 4(d). However, in supplement S5 the bulk force calculation by MEL ST is shown for a Mie or more complex objects embedded in homogeneous, heterogeneous and bounded background. As per we know, previously force calculation for heterogeneous medium has not been discussed in literature.

**Chiral modified Einstein-Laub and chiral modified Chu formulations**

In this section we shall show the consistency of the proposed internal MEL method with the external Minkowski ST method [28]. Especially for chiral objects, no built in software [39] technique is available to yield the total volumetric force. The primary goal of this section is to set an efficient computational way for embedded chiral objects. The constitutive relations inside a chiral Mie object can be written as [35]:

$$\boldsymbol{D}_{\text{in}}^{\text{chiral}} = \varepsilon_s \boldsymbol{E}_{\text{in}} - j(\kappa/c)\boldsymbol{H}_{\text{in}} ; \tag{8a}$$

$$\boldsymbol{B}_{\text{in}}^{\text{chiral}} = \mu_s \boldsymbol{H}_{\text{in}} + j(\kappa/c)\boldsymbol{E}_{\text{in}} . \tag{8b}$$

If a magneto-dielectric chiral object is embedded in a material background instead of air, the internal stress tensor that may yield the total force of the scatterer is the chiral MEL stress tensor for chiral object:

$$\left\langle \overline{\overline{T}}_{MEL(j)}^{chiral} \right\rangle (in) = D_{in}^{chiral} E_{in}^* + B_{in}^{chiral} H_{in}^* - \frac{1}{2}\left( (\frac{\mu_{b(j)}}{\mu_s})\mu_s H_{in}^* \cdot H_{in} + (\frac{\varepsilon_{b(j)}}{\varepsilon_s})\varepsilon_s E_{in}^* \cdot E_{in} \right) \overline{\overline{I}}. \qquad (9)$$

Where $j=1,2,3,...., N$ represents the number of background regions sharing interface with the chiral object. The total time averaged bulk force on the embedded object should be: $\left\langle F_{Bulk} \right\rangle (in) = \sum_j \int \left\langle \overline{\overline{T}}_{MEL(j)}^{chiral} \right\rangle (in) \cdot ds_{(j)}$. A possible derivation of Eq. (9) can be yielded very similar to our achiral MEL ST shown in supplement S3. The Bulk force of chiral MEL method can also be written as:

$$\left\langle f_{MEL(j)}^{Chiral} \right\rangle (\text{Bulk}) = \frac{1}{2}\text{Re}\left[ \left(P_{Chiral(j)} \cdot \nabla\right) E_{in}^* + \left(M_{Chiral(j)} \cdot \nabla\right) H_{in}^* - \left(i\omega P_{Chiral(j)} \times B_{in}^*\right) + \left(i\omega M_{Chiral(j)} \times D_{in}^*\right) \right]. \qquad (10)$$

A very similar formulation can also be re-written for chiral embedded objects based on the modified Chu formulation for achiral embedded objects given in [21,39]. In Eq. (10), the effective polarization and magnetization are defined as:

$$P_{chiral} = P_e + M_c, \quad P_e = (\varepsilon_S - \varepsilon_b)E_{in}, \quad M_c = -j(\kappa/c)H_{in} \qquad (11a)$$

$$M_{chiral} = M_n + P_c, \quad M_n = (\mu_S - \mu_b)H_{in}, \quad P_c = j(\kappa/c)E_{in}. \qquad (11b)$$

However, $D_{in}^{chiral}$ and $B_{in}^{chiral}$ in Eq. (10) should be written from Eq. (8a) and (8b) respectively. The surface force part has also been derived similar to achiral case discussed in previous section. The total time averaged force should be the surface force plus the bulk force. Consistency of chiral MEL ST for the unbounded homogeneous background is shown in details in supplement S6. In Fig. 5 (a)-(d), consistency of the chiral MEL ST has been shown considering a 2D infinite cylinder embedded in a heterogeneous background. Finally, the result shown in Fig. 6 [the bounded background case] bears some important physical insight. We have considered $b=800$ nm and $a=600$ nm. If $b$ is made even much smaller and very close value of $a$ (but $b>a^+$), still the conclusion presented Fig. 6 (a)-(d) remains valid. Hence we can conclude that the boundary between the scatterer and background play a vital role to yield the total time averaged force, which may not be properly explained by GAP METHOD of force calculation.

Interestingly, the internal MEL ST (internal bulk force) leads to almost time averaged total force for several situations (or at least follows the trend of total time averaged force), which can be very useful from computational point of view for both embedded chiral and achiral objects. Finally, considering all the cases discussed in this work, there is a simultaneous fulfilment, both analytically and numerically, of the 'consistency conditions' C(I) and C(II) by the external Minkowski ST and the internal modified EL or modified Chu formulation.

**CONCLUSION**

In this paper we have shown that: the 'validity condition' of $\bar{\bar{T}}(\text{in})$ and $f$ (in), defined at the beginning, plays a key role for the severe disagreement, reported in [19,38], of different force laws employed to describe real experimental results. We have proposed a solution to this problem based on the MEL ST, the MEL volume force [22] or alternatively by the modified Chu volume force [21, 39]. However, If $\langle F^{\text{Bulk}} \rangle(\text{in}) \neq 0$ for a non-absorbing object; $\bar{\bar{T}}(\text{out})$ and $\bar{\bar{T}}(\text{in})$ cannot have the same form to satisfy both consistency conditions C(I) and C(II) simultaneously, when the background is a material medium rather than air. What is the physical reason behind that? A possible answer of this important question is discussed next.

Let us consider the total momentum conservation equation [58]: $\oint \bar{\bar{T}}.ds = \int f \, dv + \frac{\partial}{\partial t}\int G \, dv$ and $p_{\text{Total}} = p_{\text{Mech.}} + p_{\text{Non-Mech.}}$ where $p$ represents momentum. Here $G$ is the electromagnetic momentum density. Though the total momentum $p_{\text{Total}}$ is always a conserved quantity, whenever one determines/measures the photon momentum transfer from the background [16,59], this leads to Minkowski's photon momentum where, rather than a simple mechanical and non-mechanical momentum part, the linear momentum equation can be better expressed as [16]: $p_{\text{Total}} = p_{\text{Cano.}}^{\text{med}} + p_{\text{Mink}}$ where $p_{\text{Mink}} = \int G_{\text{Mink}} dv$ with $G_{\text{Mink}} = D \times B$. On the other hand, Abraham photon momentum [P$_{\text{Abr}}$ (in)] is considered as the travelling momentum of the photon in a continuous medium [59], and can also be considered as the remaining electromagnetic part of the photon momentum after delivering the mechanical momentum of the photon inside an object, setting that object in motion [hence being a kinetic momentum [16]: $\boldsymbol{p}_{\text{Total}} = \boldsymbol{p}_{\text{Kin}}^{\text{med}} + \boldsymbol{p}_{\text{Abr}}$, where $\boldsymbol{p}_{\text{Abr}} = \int \boldsymbol{G}_{\text{Abr}} dv$ with $\boldsymbol{G}_{\text{Abr}} = \frac{\boldsymbol{E} \times \boldsymbol{H}}{c^2}$]. So, the role of the Minkowski and Abraham photon momenta are fully different [16, 59], and hence in the aforementioned conservation equation, $\boldsymbol{G}$ should also be different when we describe two different process: transfer of momentum from background (i.e. transfer of $\boldsymbol{p}_{\text{Mink}}$ from background to an embedded object due to the Doppler effect [16]) and delivery of momentum inside an object (i.e. $\boldsymbol{p}_{\text{Abr}}$ inside the Einstein-Balaz's box [16] according to non-relativistic analysis). This suggests that as the total momentum is a conserved quantity, $\boldsymbol{f}$ can also be different due to the different $\boldsymbol{G}$ and they should bear fully different physical meanings. Thus explanation we propose is that the time averaged Minkowski's 'total' force describes how much mechanical momentum has already been transferred to an embedded object from the background medium, which can be calculated just by employing the external fields of the embedded object. In contrast, the time averaged MEL 'bulk' force describes how much force (due to local fields) has been felt by the induced dipoles of an embedded object. Now, and most importantly, though by adding the surface force of MEL with the time averaged bulk force we get the same value of the time averaged total force as Minkowski's external force [$\langle \boldsymbol{F}_{\text{Mink}}^{\text{Total}} \rangle (\text{out}) = \langle \boldsymbol{F}_{\text{MEL}}^{\text{Bulk}} \rangle (\text{in}) + \langle \boldsymbol{F}_{\text{MEL}}^{\text{Surface}} \rangle$], they are different from a physical point of view. In fact, this issue has already been pointed in refs. [16, 59] regarding photon momenta only. Such dissimilarity may, in general, also hold for the role of the distinct time averaged stress tensors; and this has been overlooked so far. Therefore, we suggest the distinct physical meanings of the left and right hand sides of this equation: $\langle \boldsymbol{F}_{\text{Mink}}^{\text{Total}} \rangle (\text{out}) = \langle \boldsymbol{F}_{\text{MEL}}^{\text{Bulk}} \rangle (\text{in}) + \langle \boldsymbol{F}_{\text{MEL}}^{\text{Surface}} \rangle$, which is also applicable for the modified Chu formulation [21]: $\langle \boldsymbol{F}_{\text{Mink}}^{\text{Total}} \rangle (\text{out}) = \langle \boldsymbol{F}_{\text{Chu}}^{\text{Bulk}} \rangle (\text{in}) + \langle \boldsymbol{F}_{\text{Chu}}^{\text{Surface}} \rangle$.

Last but not least, our work explains not only the reason of the inconsistency of different time averaged volumetric forces reported in [19], [38], but also provides an efficient alternative solution to calculate the time-averaged bulk force with the modified Einstein-Laub (MEL) stress tensor method, saving much calculation time and memory with respect to the time-consuming bulk volumetric force method [39] for both embedded achiral [1-3,19,38-45,60] and chiral objects [26-27,32,61]. Similar time-averaged modified formulations are also possible on employing modified Chu methods [21,39].

In fact, the main goal of this paper is not to show superiority of one formula over another. Rather the main targets of this work were to make fully distinct formulas mathematically equivalent to yield accurate and consistent time-averaged total optical force, as well as explaining the exact reason of discrepancies of previous theories for real experimental observations.

ACKNOWLEDGEMENTS

MN-V acknowledges MINECO, grants FIS2012-36113-C03-03, FIS2014-55563-REDC and FIS2015-69295-C3-1-P. W.D. acknowledges National Natural Science Foundation of China under grant number 11474077. M.R.C.M. acknowledges Associate Professor Qiu Cheng Wei in National University of Singapore (NUS) for important discussions throughout this work. We also acknowledge Dr. Maoyan Wang in University of Electronic Science and Technology of China.

**Figures and Captions**

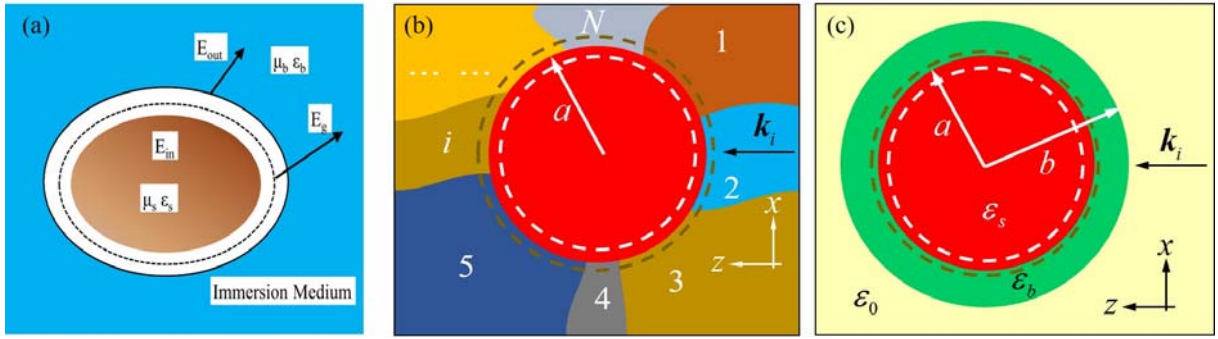

Fig. 1 Procedure of time-averaged optical force calculation by employing stress tensors. (a) GAP METHOD: a small gap between the scatterer and the background should be considered. The time-averaged total force on the scatterer should be calculated using the time-averaged ST. $\langle \boldsymbol{F}_{\text{out}}^{\text{GAP}} \rangle$ evaluated *from fields strictly outside the object [i.e. gap field and background field]*, putting the integration boundary in the gap [20] (black dashed circle). However, the volume force calculation method is a little bit different, which is discussed in detail in [23]. (b) and (c) NO GAP METHOD: In both examples the total force obtained by using the time-averaged ST is $\langle \boldsymbol{F}_{\text{out}} \rangle$ evaluated *from fields strictly outside the object considering no gap*, at $r = a^{+} = 1.001a$, (black circles); whereas this force is $\langle \boldsymbol{F}_{\text{in}} \rangle$ or bulk force when the ST is determined *from fields strictly inside the object* considering no gap at $r = a^{-} = 0.999a$ (white circles). In (b), a sphere or cylinder is immersed in an unbounded and heterogeneous background. In (c) a core-shell sphere or cylinder (i.e., the core is embedded in a bounded background).

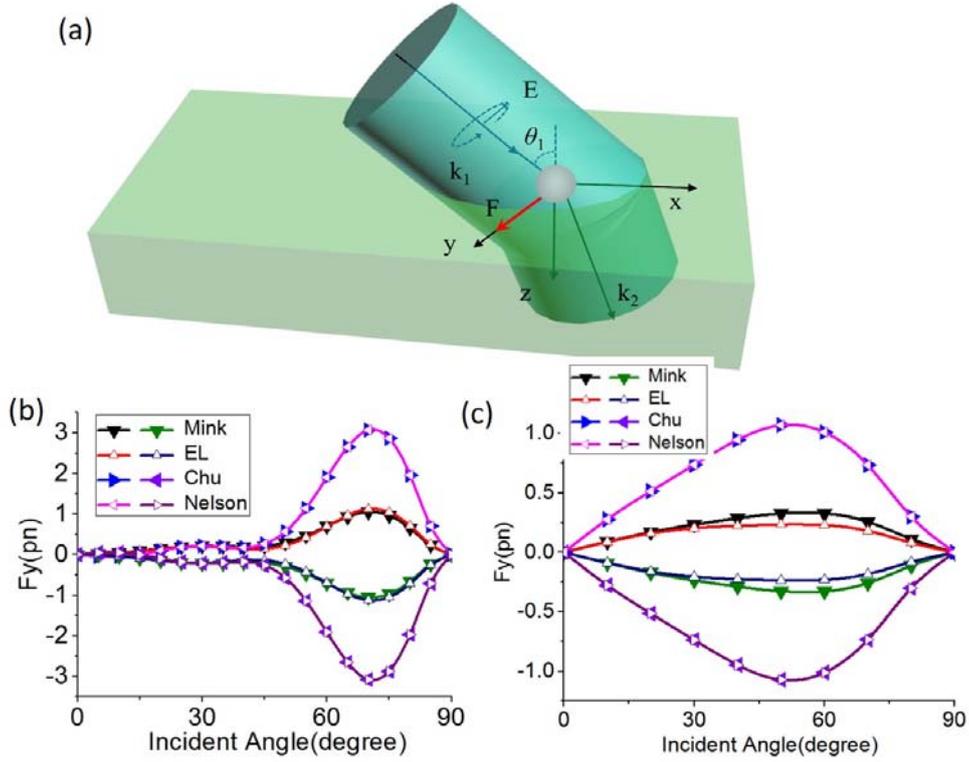

Fig. 2 (a) 3D Illumination geometry for a particle at the interface between two media to model the lateral force set-up in [12]. Either a right-handed or a left-handed circularly polarized (CP) beam of wavelength 1064 nm is incident along the x–z plane at angle $\theta_1$. $\mathbf{k_1}$ and $\mathbf{k_2}$ are wave vectors in media 1 and 2, respectively. The external force has been calculated based on the NO GAP METHOD. (b) Transversal force (negative for a left- handed CP and positive for right-handed CP) as a function of angle of incidence for a 500 nm (radius) spherical $TiO_2$ particle located at the water–air interface calculated by external Minkowski, Einstein–Laub, Chu, and Amperian/Nelson ST. The Minkowski and Einstein-Laub STs predict a smaller lateral force in comparison with the time-averaged force yield by the external Chu and Ampere ST. (c) Transversal force, (negative for left-handed CP and positive for right-handed CP) as a function of the angle of incidence for an elliptical polystyrene (PS) particle [$r_x$=800 nm and $r_y$=$r_z$=(800/3) nm] located at the water–air interface, calculated by external Minkowski, Einstein–Laub, Chu, and Amperian STs. The Minkowski and Einstein-Laub STs predict lower lateral force in comparison with the time-averaged force yielded by the external Chu and Ampere STs.

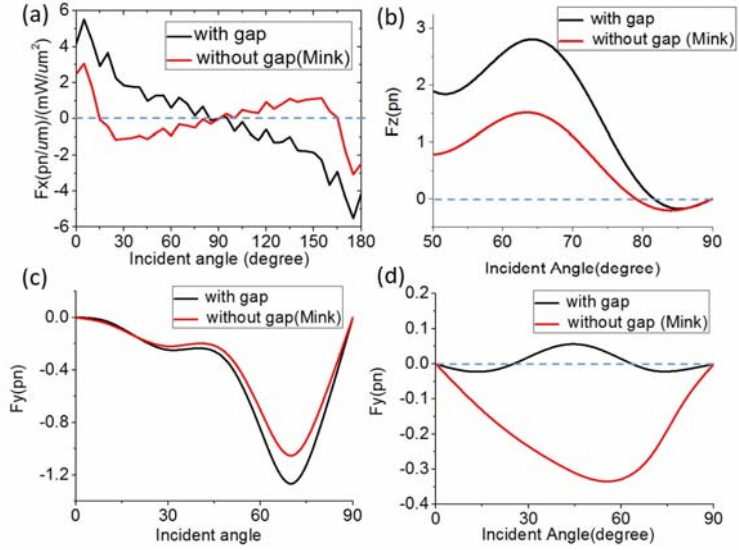

Fig. 3 Illustration of the optical force obtained from the GAP METHOD. (a) Optical force for a 2D spherical scatterer placed in an air-water interface, (cf. Fig. 2(d) and Fig. 2(e) in [3]). The parameters are $n_1$=1.0, $n_2$=1.33, $n_3$=1.45. The variation in the optical forces with the incidence angle [3] for the p-polarization case, as calculated via the external vacuum stress tensor considering a small gap of 2 nm, [the possible smallest gap with 2D full wave simulation set-up; gap size << incident wavelength], between the scatterer and the water background. We have also examined our results by varying the size of the artificial gap, (i.e. 6nm, 10 nm and 20 nm). The results are almost same for all those gaps. The size of the scatterer is defined by $r_x = r_y$ =2.0 μm [3]. Instead of optical pulling [3], an optical pushing is achieved from the GAP METHOD. (b) Optical force (obtained by the vacuum ST) of a 3D dielectric particle given in Supplement S1, Fig. 1s (c), using two obliquely incident plane waves *but considering a small gap of 10 nm* [possible smallest gap with a 3D full-wave simulation set-up; gap size << incident wavelength] between the water background and the embedded scatterer. Both GAP and NO GAP METHODS lead to consistent results. (c) For the 3D set-up of Fig. 2(b), the time-averaged lateral force has been calculated from the vacuum ST considering a small gap (10nm) between the scatterer and the water background for a left-handed CP wave. (d) For the 3D set-up of Fig. 2(c), the time-averaged lateral force has been calculated from the vacuum ST considering a small gap (10 nm) between the scatterer and the water background for a left-handed CP wave. The sign of the lateral force is in disagreement with the time- averaged force yielded by the NO GAP METHOD in Fig. 2(c).

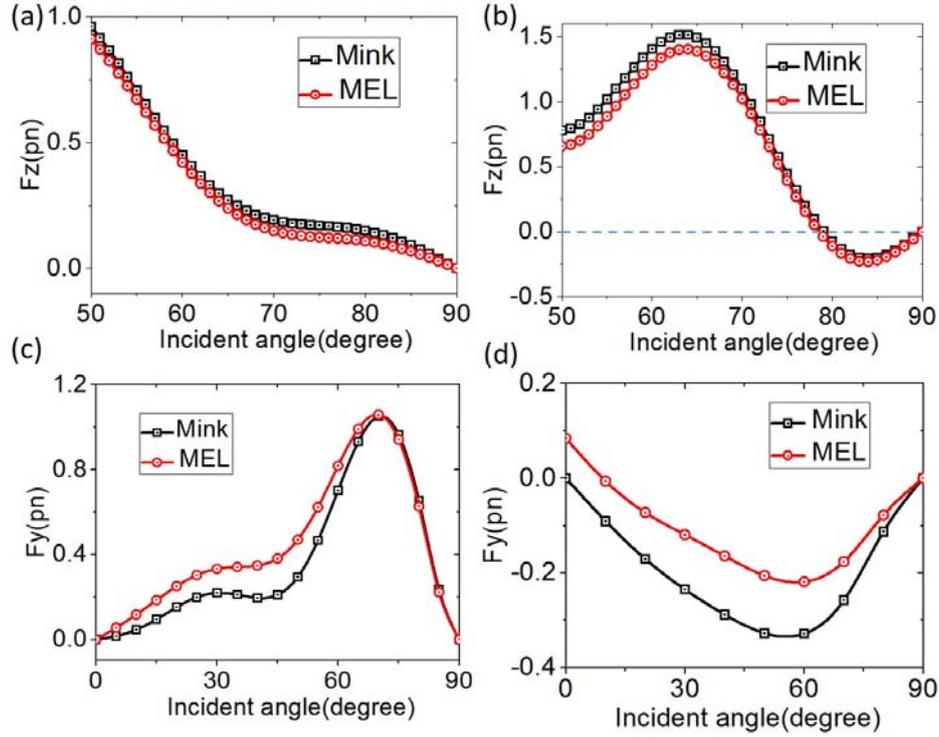

Fig. 4 Calculation of time averaged total optical force (NO GAP METHOD) by external Minkowski ST and the time averaged bulk force by internal MEL ST. These forces are always of same trend. By adding the surface force of achiral MEL with bulk force [cf. Eqs (6) and (7)], the magnitude exactly matches with external time averaged total force by Minkowski ST. (a) For the two beam tractor set-up in Supplement S1 Fig. 1s(b) with 320 nm object. (b) For the two beam tractor set-up in Supplement S1 Fig. 1s(c) with 410 nm object. (c) For the lateral force set-up in Fig. 2(b) [only left hand CP wave incident case] with spherical object. (d) For the lateral force set-up in Fig. 2(c) [only right hand CP case] with elliptical object. For all the cases the trend of the time-averaged bulk force, obtained by employing the internal field only, is very similar to the total outside force calculated by the external Minkowski ST, using fields exterior to the scatterer. The bulk force by modified Chu volume force or stress tensor does not follow the trend of the total force. Moreover, the internal force calculation by the MEL ST is very less-time and memory consuming in comparison with the modified volumetric force calculation method [cf. the detailed discussion in [21] and [39]]. These are the main computational advantages of MEL ST method.

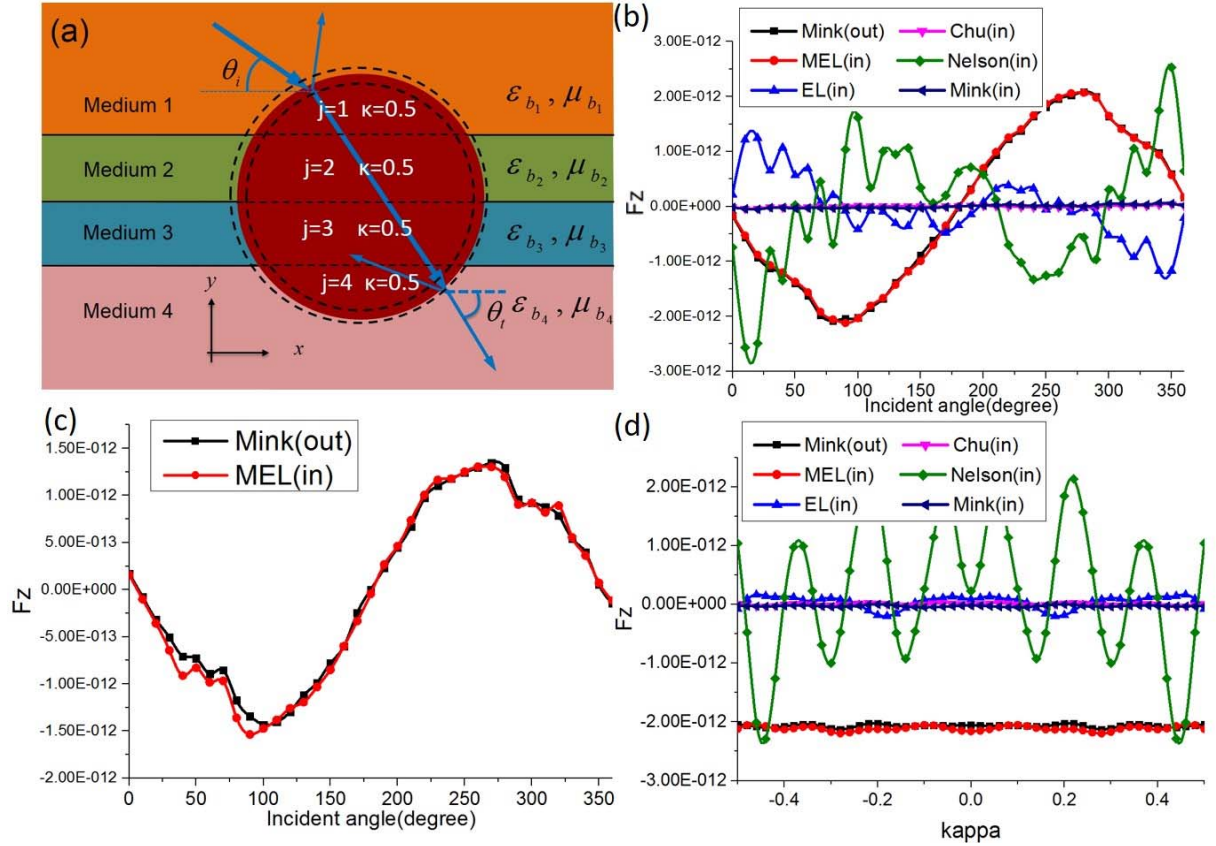

FIG. 5 Time-averaged forces: $F_{out}$ at $r = a^+ = 1.001a$ from Minkowski ST and $F_{in}$ (bulk force) at $r = a^- = 0.999a$ from the Chiral MEL ST. These forces have always the same trend. By adding the surface force of Chiral MEL with bulk force, the magnitude exactly matches with the external time-averaged total force. (a) Calculation procedure of force on a magneto-dielectric infinite chiral cylinder (chirality parameter, $\kappa = 0.5$) of $(\varepsilon_s, \mu_s) = (5\varepsilon_0, 4\mu_0)$ and radius 2000 nm, embedded in an heterogeneous unbounded background of four different magneto-dielectric layers: $(\varepsilon_b, \mu_b) = (3\varepsilon_0, 2\mu_0)$; $(4\varepsilon_0, 3\mu_0)$; $(5\varepsilon_0, 4\mu_0)$; $(6\varepsilon_0, 5\mu_0)$ at $\lambda = 1064$ nm. (b) Force on that cylinder when the plane wave $E_x = E_0 e^{i(kz-\omega t)}$ illuminates at varying angles of incidence. Notice that the internal force calculated by all other STs (i.e. EL, Chu, Nelson and Minkowski) has not in the same trend as the total external force. (c) Force on the same embedded cylinder when the illuminating circularly polarized wave: $E_x + iE_y : E_x = E_0 e^{i(kz-\omega t)} = E_y$ incides at varying angles of incidence. (d) Force on the cylinder versus varying chiral parameter $\kappa$. when the illuminating plane wave incides at an angle of 45 degrees.

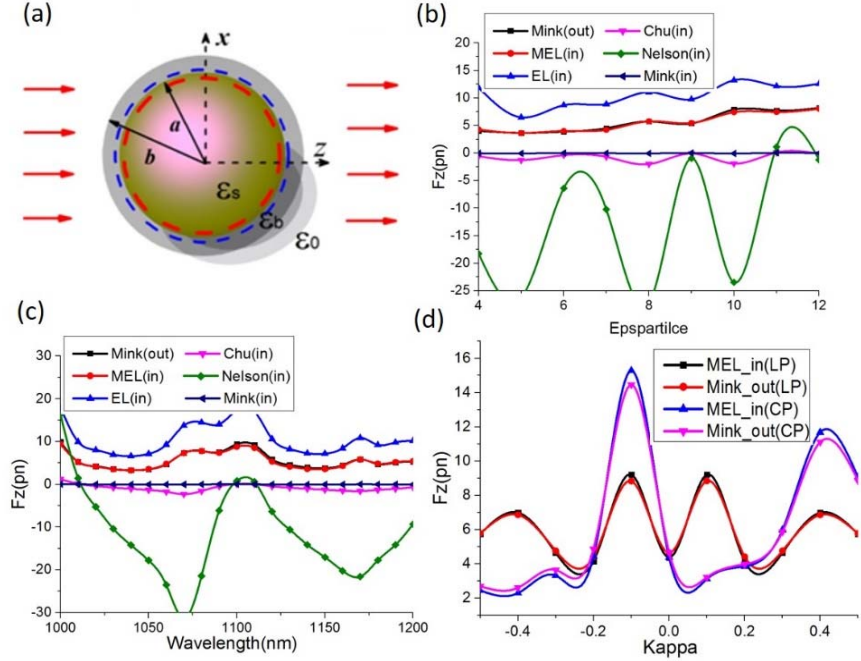

FIG. 6: Time-averaged forces: $F_{out}$ at $r=a^+ =1.001a$ from Minkowski ST and $F_{in}$ (bulk force) at $r=a^- =0.999a$ from the Chiral MEL ST. These forces are always of same trend. By adding the surface force of Chiral MEL with bulk force, the magnitude exactly matches with the external time-averaged total force. (a) Calculation procedure of a 3D magneto-dielectric core where the whole core-shell sphere is embedded in air. Core radius, $a$=600 nm, $\varepsilon_s = 8\varepsilon_0$, $\mu_s = 4\mu_0$ Bounded local immediate background (i.e. the shell) parameters: radius, $b$=800 nm and $\varepsilon_s = 4\varepsilon_0$; $\mu_s = 2\mu_0$. This entire core-shell is illuminated at a wavelength of $1070$ nm. (b) For plane wave illumination, $E_x = E_0 e^{i(kz-\omega t)}$: $\langle F_{out}^{Core} \rangle$ at different chirality parameters obtained from Minkowski ST at $r=a^+$ using the fields in the shell. Force $\langle F_{in}^{Core} \rangle$ based on the Chiral MEL ST at $r=a^-$ using core fields. The bulk force on the core given by other STs do not follow the trend of the total external force. (c) For circularly polarized wave illumination ($E_x + iE_y : E_x = E_0 e^{i(kz-\omega t)} = E_y$): still the bulk force by Chiral MEL ST is of the same trend as the external time-averaged total force as the chirality parameter of the core varies. (d) For linear and circularly polarized wave illumination: again our conclusions remain valid as the chirality parameter $\kappa$ of the core varies.